\documentclass[11pt,letterpaper]{article}
\usepackage{amsmath,cite,amssymb,color,feynmf}

\definecolor{darkred}{rgb}{0.6,0.0,0.0}
\definecolor{darkblue}{rgb}{0.0,0.0,0.7}
\definecolor{darkgreen}{rgb}{0.0,0.5,0.0}

\def\rad{\mathcal{T}}

\newcommand{\beq}{\begin{equation}}
\newcommand{\eeq}{\end{equation}}
\newcommand{\beqa}{\begin{eqnarray}}
\newcommand{\eeqa}{\end{eqnarray}}

\newif\ifdohyperref
\dohyperreffalse

\ifdohyperref
  \usepackage{hyperref}
  \hypersetup{backref}
  
  \newcommand\eprint[1]{\href{http://arXiv.org/abs/#1}{[arXiv:#1]}}
\else
  
  \newcommand\eprint[1]{[arXiv:#1]}
\fi

\begin{document}

\begin{titlepage}
\begin{flushright}
\end{flushright}

\vspace{15pt}

\begin{center}

{\Huge\bf  Supersymmetry Breaking and\\ Moduli Stabilization in AdS}

\vspace{15pt}

{\Large{Andrey Katz}$^a$\footnote{andrey@physics.technion.ac.il},
\Large{Michele Redi}$^b$\footnote{redi@physics.nyu.edu}, \Large{Yael
Shadmi}$^a$\footnote{yshadmi@physics.technion.ac.il} and \Large{Yuri
Shirman}$^c$\footnote{shirman@lanl.gov} }

\vspace{7pt}
{\small
${}^a$
{\it{Physics Department, Technion---Israel Institute of Technology,
Haifa 32000, Israel \\} }
\vspace{4pt} ${}^b$
{\it{CCPP, Department of Physics, New York University, 10003, New York, NY\\} }
\vspace{4pt}
${}^c$ {\it{T-8, MS B285, LANL, Los Alamos, NM 87545}
}
}
\end{center}

\begin{abstract}
We study the one loop effective potential for the radion superfield
in the supersymmetric Randall-Sundrum scenario with detuned brane
tensions. At the classical level the distance between the branes is
stabilized while the VEV of the fifth component of the graviphoton
is a flat direction which breaks supersymmetry. At the quantum level
a potential is generated. This leads to a toy model of a
supersymmetric compactification with all the moduli stabilized
perturbatively.
\end{abstract}

\end{titlepage}

\section{The Model}
One of the main problems facing compactifications of
higher-dimensional theories down to four dimensions, is the
stabilization of all the moduli of the compactification. In the
context of string theory, a possible solution to this problem has
been proposed recently where classical effects, such as fluxes,
combine with non-perturbative effects to give a potential for all
the moduli in a supersymmetric AdS vacuum \cite{kklt}. By adding
supersymmetry breaking effects, it was argued that the cosmological
constant could also be lifted to a small positive value, realizing
the necessary starting point for phenomenological applications.

The purpose of this note is to study a mechanism of moduli
stabilization similar to \cite{kklt}, where \emph{perturbative} and
fully \emph{calculable} corrections stabilize all the moduli of the
compactification. The model in question is the supersymmetric
Randall-Sundrum (RS) scenario with detuned brane tensions
\cite{Randall:1999ee,old,rk,detuned}. In the minimal scenario the
bosonic part of the action contains, beside the graviton, a $U(1)$
gauge field $B_M$ called the graviphoton. In the $4D$ effective
theory the radion is accompanied by an axion arising from the fifth
component of the graviphoton field, and together they form the
complex scalar of a $4D$ chiral multiplet. Already at the classical
level the radion has a potential with mass proportional to the
$4D$ curvature. The axion partner on the other hand remains a flat
direction of the potential as a consequence of the higher
dimensional gauge invariance. It was however noticed in
\cite{Bagger:2003fy,Bagger:2003dy} that supersymmetry would be
broken spontaneously by a non-zero vacuum expectation value (VEV) of
the graviphoton modulus. This is due to the fact that the
graviphoton gauges a $U(1)$ $R-$symmetry of the $5D$ action, under
which the gravitino is charged. The gauging violates explicitly the
shift symmetry of $B_5$ and upon compactification contributes to the
gravitino bilinears which in turn break supersymmetry by shifting
the masses of the fermionic Kaluza-Klein (KK) tower.

With broken supersymmetry, the potential is modified by radiative
corrections and the flat directions are lifted, because the fermion
and boson contributions do not cancel exactly\footnote{This
mechanism is similar to the one considered in \cite{quiros} in flat
space with the important difference that the supersymmetry breaking
parameter is a dynamical variable.}. In principle the effective
potential could be found by brute force, by computing the
Coleman-Weinberg-like potential obtained by integrating out the full
KK tower of modes. This computation is highly unwieldy mostly
because the $4D$ ground state is curved and the masses of the modes
are not known in closed form (see \cite{norman}). An interesting
feature however is that, since the supersymmetry breaking effect is
non-local in the extra-dimension, the final result is guaranteed to
be finite and calculable, despite the appearance of divergences at
each KK level. In this paper we will follow a different path and
rely on the power of supersymmetry to derive the effective
potential. Since the superpotential is not renormalized
perturbatively all that is needed is the correction to the K\"ahler
function. As we will argue this allows an enormous simplification
because it is sufficient to compute the correction in the limit
where the $4D$ ground state is flat and supersymmetry is preserved.
As expected, a potential for the graviphoton modulus is generated.
The correction to the potential is negative with pure gravity in the
bulk, so that unbroken supersymmetry corresponds to a global maximum
of the potential while the minimum of the potential corresponds to
maximal supersymmetry breaking. Nevertheless, because the ground
state is AdS, unbroken supersymmetry remains a stable point of the
potential. With the addition of hyper-multiplets in the bulk, the
correction can become positive, so that unbroken supersymmetry is
the minimum of the potential.

While we focus on a specific model we would like to emphasize that
the mechanism presented in this paper could be generic in
supersymmetric compactifications. In \emph{any} supersymmetric model
where the ground state is AdS space, half of the moduli have  mass
simply because the scalars in a chiral multiplet have different
masses (split by the curvature of the space). Very generically in a
supersymmetric compactification the imaginary components of the
chiral fields are axions arising from the form-fields of the higher
dimensional theory. The axions are flat directions at tree level due
to the shift symmetry inherited from the gauge invariance of the
theory. If the flat direction breaks supersymmetry as in the case
considered in this paper, the masses of all the moduli will be
lifted. The examples that we have in mind are gauged supergravities
where the shift symmetry of the axions arising from the gauge fields
is violated by the gauging. In this type of models, without invoking
non-perturbative contributions, one expects that in an AdS
supersymmetric vacuum all the moduli will acquire a mass
proportional to the four dimensional curvature. If the minimum of
the potential can be lifted without affecting significantly the
stabilization (which might not be generic), this would lead to a
very peculiar spectrum of masses where half of the scalars have mass
at least a loop-factor smaller than the one of the scalar partner.

\section{Gravitational Multiplet}

The $4D$ low energy effective theory for the supersymmetric RS model
with general brane tensions was computed in \cite{Bagger:2003dy}, at
the classical level. The low energy dynamics is described by an
$N=1$ supersymmetric sigma-model coupled to supergravity, with the
following K\"ahler potential and superpotential,
\begin{eqnarray}
&&K(\rad,\bar{\rad}) = -3 M_4^2 \log\left(1 - e^{-k \pi (\rad+\bar{\rad})}
\right)\  \nonumber \\
&&W(\rad)= \sqrt{1-e^{-2k\pi r_0}}\, \frac {M_4^2}{L}  \left(1 -e^{i
\phi} e^{\pi k r_0}  e^{-3 \pi k \rad}\right)\ , \label{super}
\end{eqnarray}
to leading order in the $4D$ curvature (see discussion below).
Here $\rad$ is the radion superfield, whose scalar component is $r+
i b$, where $r$ is the radion and $b$ is the zero mode of $B_5$. The
parameterization of $K$ and $W$ is chosen so that the minimum of the
potential is located at $r=r_0$, and $L$ is the radius of the $4D$
AdS ground state while $k$ is the curvature of the bulk AdS$_5$.

An interesting feature of the low energy effective
action~(\ref{super}), is that it is entirely determined by the shift
symmetry of $b$ up to the phase $\phi$ which can be set to zero
without loss of generality (it amounts to changing the origin of
$b$) \cite{Bagger:2003dy}. Since the bosonic part of the action is
invariant under the shift of $B_5$, there is no potential for $b$ at
tree level\footnote{This symmetry is indeed violated by the
Chern-Simons term but this is irrelevant in perturbation theory.}.
This condition, together with the fact that the ground state is AdS,
fixes the form of $K$ and $W$ as in (\ref{super}). Even though there
is no potential for $b$ at tree level, the physics does depend on
its VEV, because the shift symmetry is violated in the fermionic
sector by the superpotential. In particular, computing the covariant
derivative $D_{\rad} W$, one can see that supersymmetry is broken
unless $b=2n/(3k)$. All this has a beautiful explanation in terms of
the dual CFT description of the model \cite{cft} (see also
\cite{Gregoire:2004nn}). From the holographic point of view, the radion
is the Goldstone boson for the spontaneous breaking of conformal
invariance. In fact the K\"ahler potential and superpotential are
dictated by conformal invariance\footnote{The only potential for the
radion compatible with conformal invariance is $\phi^4$ which is
precisely what follows (in the rigid limit and upon canonical
normalization) from the exponential term in $W$.}. The constant
piece in $W$ arises due to the explicit breaking of conformal
invariance in the ultraviolet which also induces the coupling to
$4D$ gravity. It is very useful to consider the physics of $b$ in
the CFT picture. The photon gauges a $U(1)$ subgroup of the
$R-$symmetry which is broken by the boundary conditions. According
to the AdS/CFT dictionary, gauge symmetries on the AdS side are dual
to global symmetries of the CFT, so $b$ is the Goldstone boson for
the breaking of the $U(1)$ $R-$symmetry of the CFT. Being a
Goldstone boson there cannot be any potential. The constant term in
the superpotential however breaks the $R-$symmetry and therefore a
potential is generated at one loop.

Let us now turn to the explicit computation of the quantum effects.
To any order in perturbation theory the superpotential is not
renormalized and is given by the tree-level result above. Since the
fermionic part of the action does not respect the shift symmetry of
$b$, at one loop there will be corrections to $K$ which do not
respect the structure of eq.~(\ref{super}) and generate a potential
for $b$. The effective action is an expansion in derivatives and in
powers of the curvature $1/L^2$. As explained in detail in
\cite{Bagger:2003dy}, working consistently to two derivatives
requires that we also work to leading order in $1/L^2$, because
terms such as $R^n$ are not included. This observation simplifies
the computation enormously. Since the superpotential is already of
order $1/L$, it follows that to work to two derivatives we just need
to compute the K\"ahler potential to zero order in $1/L$. What this
means is that to calculate the one loop potential, all we need is
the correction to the K\"ahler potential in the supersymmetric RS
model with tuned brane tensions. In fact, the very same arguments
could be used for any theory with unbroken supersymmetry in AdS
background.

In the flat supersymmetric limit the shift of $b$ becomes an exact
symmetry so that $K$ is a function of $\rad+\bar{\rad}$. The
correction can be derived from the remarkably simple formula
\cite{Gregoire:2004nn},
\begin{equation}
\Delta \Omega_{gravity}= \sum_n \int \frac {d^4 p}{(2\pi)^4} \frac 2
{p^2} \log(p^2+m_n^2)
\end{equation}
where the sum runs over the masses of the KK tower and
$\Omega=\Omega_{tree}+\Delta\Omega$ is related to the K\"ahler
potential by $K=-3M_4^2 \log[k\, \Omega/(3 M_5^3)]$ where $M_5$ is
the $5D$ Planck mass (we follow the normalizations in
\cite{Bagger:2003dy}). Using dimensional regularization one finds,
\begin{equation}\label{dimregsum}
\Delta \Omega_{gravity}= - 2 \frac {\Gamma(1-d/2)} {(4\pi)^{d/2}}
\sum_n m_n^{d-2}\ ,
\end{equation}
for the gravity multiplet. The finite, radion dependent contribution
arising from the sum~(\ref{dimregsum}) can then be rewritten as
\cite{goldberger,Falkowski:2005fm},
\begin{eqnarray}
\Delta \Omega_{gravity}&=&\frac {k^2 a_\pi^2} {4 \pi^2}\,
\int_0^\infty dy~ y \log \left(1-\frac {I_1(y a_\pi) K_1(y)}{K_1(y
a_\pi) I_1(y)}\right)\nonumber \ ,
\\  a_\pi^2&=&e^{-k \pi
(\rad+\bar{\rad})}\ . \label{omegagr}
\end{eqnarray}

The above result is exact at one loop for any value of the bulk
curvature $k$. The formulas simplify when the warping is large so
we focus on this case. In this limit one
obtains~\cite{Gregoire:2004nn,Falkowski:2005fm},
\begin{equation}
\label{kloop} K=-3 M_4^2 \,\log \left[(1+\alpha)-(1+\beta) e^{-k \pi
(\rad+\bar{\rad})} +\gamma e^{-2 k\pi
(\rad+\bar{\rad})}+\dots\right]\, ,
\end{equation}
where the dots stand for higher terms in the exponential expansion.
The last term is the finite calculable contribution obtained from
(\ref{omegagr}). The precise numerical coefficient is given by,
\begin{eqnarray}
\gamma=-\frac {c_G\, k^2}{12 \pi^2 M_4^2}\ ,
\end{eqnarray}
where $c_G=\frac12\, \int_0^\infty dx\, x^3 \,
\frac{K_1(x)}{I_1(x)}
 \approx 1.165$.
In this formula we have used the relation
$M_4^2=M_5^3/k$ which is valid to leading order in the exponential
expansion.
This term does not respect the tree level structure of $K$ so as we
will show it gives rise to a $b$ dependent potential.

The parameters $\alpha$ and $\beta$ parameterize renormalizations of
tree-level terms. Unlike $\gamma$ they do not generate a
$b$-dependent potential. If we are only interested in the $b$
dependent correction to the potential, we could just ignore the
corrections $\alpha$ and $\beta$ which will only enter at two-loops.
Still, it is instructive to consider these divergent parameters. As
shown in~\cite{Gregoire:2004nn}, $\alpha$ and $\beta$ correspond to
divergent brane terms generated radiatively on the ultraviolet
($UV$) and infrared ($IR$) branes respectively. In particular, the
brane terms contain the Ricci scalar, which has a non-zero VEV in
AdS, and therefore contributes to the effective brane tensions.
Hence $\alpha$ corresponds to a correction of the $UV$ brane
tension. Since the detuning of the $UV$ brane compared to the bulk
cosmological constant determines the $4D$ curvature $1/L$ (we use
the parameterization of \cite{Bagger:2003dy}), a non-zero $\alpha$
modifies the value of $L$. Indeed, from the point of view of the
$4D$ theory, a non-zero $\alpha$ in~(\ref{kloop}) changes the
overall scale of the potential, and therefore shifts the value of
the $4D$ curvature $1/L$ with respect to the $4D$ Planck scale.
Similarly, $\beta$ corrects the $IR$ brane tension, which determines
the radius. Thus the only effect of the divergent parameters
$\alpha$ and $\beta$ is to modify the radius and curvature of the
$4D$ theory.

From the discussion above it follows that $\alpha$ and $\beta$
should be fixed by matching to the $5D$ theory. Consider the
contribution of some KK supermultiplet to the vacuum energy. With
unbroken supersymmetry the contribution vanishes identically.
Therefore, a natural matching condition is that the one-loop
correction to the potential vanishes when supersymmetry is unbroken,
\begin{equation}
\Delta V(r,b)=0 \ \ {\mathrm{for}}\ \ r, b \ {\mathrm{such\ that}}\
\ \ D_\rad W=0\ ,
\end{equation}
where $\Delta V$ is the one-loop correction to the potential
obtained from the K\"ahler potential~(\ref{kloop}) and
superpotential~(\ref{super}) through the standard supergravity
formula,
\begin{equation}
V=e^{\frac {K} {M_4^2}} \left(K^{\rad\bar{\rad}}D_\rad W {D_{\bar
\rad} \bar{W}}- \frac 3 {M_4^2} W\bar{W}\right). \label{sugrapot}
\end{equation}
The second condition that we require is that the value of the radion
at the minimum does not change when supersymmetry is unbroken. These
two conditions determine,
\begin{eqnarray}
\alpha&=&\gamma\,  e^{-4 k\pi r_0}\ ,\nonumber \\
\beta&=&2\gamma\, e^{-2k \pi r_0}\,.
\end{eqnarray}
to leading order in $e^{-2 k \pi r_0}$.

Given $K$ and $W$ we can now compute the potential from
(\ref{sugrapot}).  In the large warping limit we find,
\begin{eqnarray}
\delta V= &-&\frac{c_G}{4\pi^2} { (e^{-2k\pi r_0} k)^2\over L^2}\,
\Bigg[ 3 - 4\, e^{-2\pi k(r-r_0)} +e^{-4\pi k (r-r_0)}\nonumber \\
&-&4\,e^{-5\pi k(r-r_0)} + 4\,e^{-6\pi k(r-r_0)} + 8\,e^{-5\pi
k(r-r_0)}\, \sin^2\left({3\over2} \pi k b\right)\Bigg]~.
\label{corr}
\end{eqnarray}

Using the fact that the mass of the first KK mode is roughly $\pi k
e^{-k \pi r_0}$, one can write the $b$-dependent piece in the more
suggestive form,
\begin{equation}
\label{gravcor} \delta V=  - \frac {2c_G}{ \pi^4} \frac
{m_{KK}^2}{L^2} e^{-2\pi kr_0} \sin^2\left(\frac {3 k \pi b}
2\right) \ ,
\end{equation}
with $m_{KK}\equiv\pi k e^{-k \pi r_0}$.
This is essentially the result that one would guess from effective
field theory considerations alone. Since the supersymmetry breaking
scale obtained from (\ref{super}) is proportional to $1/L$ and the
low energy effective theory is cut-off at the KK scale $m_{KK}$, the
correction computed within the effective theory must be proportional
to $m_{KK}^2/L^2$. The extra suppression is related to the fact that
the supersymmetry breaking parameter is suppressed in comparison to
$1/L$. This can be seen from (\ref{super}) because even for maximal
supersymmetry breaking the shift of the gravitino mass is only,
\begin{equation}
\delta m_{\frac 3 2}\approx\, \frac {e^{-2 k \pi r_0}} L\ .
\end{equation}

The correction~(\ref{corr}) is always negative, so the
supersymmetric vacuum $b=0$ is a maximum of the potential. The
global minimum corresponds to maximal supersymmetry breaking. At the
supersymmetric stable point the mass of $b$ is given by,
\begin{equation}
m^2_b=-\frac{3c_G}{2\pi^4}\frac{1}{L^2}\frac{m_{KK}^2}{M_4^2}\ .
\label{massbsw}
\end{equation}
Even though $b=0$ is a maximum of the potential, it remains a stable
point by virtue of the fact that the ground state is AdS. In other
words, radiative corrections cannot destabilize the supersymmetric
vacuum. In fact supersymmetry automatically guarantees that all the
masses are above the stability bound for the scalars in AdS$_4$,
$m^2\ge -9/(4L^2)$. The radion mass is also corrected as,
\begin{equation}
\label{mr} m_r^2= \frac 4 {L^2}
-\frac{5c_G}{2\pi^4}\frac{1}{L^2}\frac{m_{KK}^2}{M_4^2}\ .
\end{equation}
Together the masses of the scalars correspond to a supersymmetric
multiplet labeled by the Casimir,
\begin{equation}
E=\frac 3
{L^2}-\frac{c_G}{2\pi^4}\frac{1}{L^2}\frac{m_{KK}^2}{M_4^2}\ .
\end{equation}

\section{Vector and Hyper Multiplets}

In the presence of vector and hyper multiplets in the bulk there
will be additional corrections to the quantum potential. While the
contribution of vector multiplets is expected to have the same sign
as the gravity multiplet, hyper-multiplets should give the opposite
sign. To see this, recall that from the five-dimensional point of
view, the physics of these corrections is the following: the action
in flat space is invariant when $B_5$ shifts by a constant. The AdS
theory is obtained by gauging a $U(1)$ subgroup of the $SU(2)$
$R-$symmetry. This breaks explicitly the shift symmetry so one
expects perturbative corrections to the potential for $B_5$. The
fields transforming under the $R-$symmetry are the gravitinos, the
gauginos, and the hyper-scalars. Since a VEV of $B_5$ shifts the
fermion masses for the gravity and vector multiplets, and the boson
masses for the hypers, they will contribute with opposite sign.

The full supersymmetric $5D$ action coupled to branes has not yet
been constructed. Still, as in the gravity case, we can circumvent this
obstruction by computing the correction to the K\"ahler potential
since this can be obtained in the tuned limit and depends only on
the KK spectrum. The formula (\ref{omegagr}) generalizes to,
\begin{equation}
\Delta \Omega=N \frac {k^2 a_\pi^2} {8 \pi^2}\int_0^\infty dy\, y
\log Z(y)\
\end{equation}
where $N$ is the number of vector multiplets $N_V$, or minus the
number of hypermultiplets $N_H$ respectively. $Z(y)$ is a function
whose zeros on the positive imaginary axis are the masses of the KK
particles which is given respectively by \cite{pomarol},
\begin{eqnarray}
Z(y)_{hyper}&=&1-\frac {I_{|c+1/2|}(y a_\pi) K_{|c+1/2|}(y)}
{K_{|c+1/2|}(y a_\pi) I_{|c+1/2|}(y)}\nonumber \\
Z(y)_{vector}&=&1-\frac {I_0(y a_\pi) K_0(y)} {K_0(y a_\pi)
I_0(y)}\, . \label{kkmasses}
\end{eqnarray}
The contribution of the hyper-multiplets depends on the parameter
$c$ which is related to the bulk mass. Since for any $c$ the KK
reduction of a hyper-multiplet always produces a massless chiral
multiplet and a tower with masses starting at $\pi k e^{-k \pi
r_0}$, the hyper-multiplets do not decouple for large mass. What
depends on $c$ is instead the localization of the zero mode.
Expanding the result in the limit of large warping one finds,
\begin{equation}
\Delta \Omega_{hyper} = N_H\, \frac {c_H} {8 \pi^2}\, k^2
e^{-(|c+{1\over2}|+1) \, k\pi (\rad+\bar\rad)}\ ,
\end{equation}
with the numerical coefficient given by
\begin{equation}
c_H=
\frac{2^{1-|2c+1|}}{\Gamma(\vert c+\frac 1 2\vert)
\Gamma(1+\vert c+\frac 1 2 \vert)}\,\nonumber\\
\int_0^\infty dy\, y^{2\vert c +{\frac 1 2}\vert +1} \frac
{K_{|c+1/2|}(y)}{I_{|c+1/2|}(y)}\ .
\end{equation}
Note that the functional dependence of the correction to $\Omega$
depends on $c$, that is on the localization of the zero mode. For
the special value $c=1/2$, the ``conformal hypermultiplet'', the
contribution is minus a half the one of gravity. The potential for
$N_H$ such multiplets is then,
\begin{equation}
\label{gravcorh} \delta V=  \left( \frac {N_H} 2 -1\right) \frac
{2c_G}{ \pi^4} \frac {m_{KK}^2}{L^2} e^{-2\pi kr_0}
\sin^2\left(\frac {3 k \pi b} 2\right) \ .
\end{equation}
For $N_H> 2$ the correction is positive and the unbroken
supersymmetry point $b=0$ becomes the minimum of the potential.

For completeness the result for the vector multiplet is given by,
\begin{equation}
\Delta \Omega_{vector} = -N_V\, \frac {c_V} {8 \pi^2}\, \frac k
{\pi(\rad+\bar{\rad})} e^{-k \pi (\rad+\bar{\rad})}\ ,
\end{equation}
with the numerical coefficient $c_V =\int_0^\infty dx\, x
K_0(x)/I_0(x) \approx .631$.

\section{Outlook}

In this brief note we computed the one loop effective potential for
the radion superfield in the supersymmetric detuned RS model. At
tree level, the zero mode of $B_5$ is an exactly flat direction of
the potential, with supersymmetry spontaneously broken for $b\neq
0$. The scalar partner of $b$, the radion, is already stabilized at
tree level. Due to supersymmetry breaking effects, $b$ develops a
periodic potential at one-loop. This potential is finite because the
supersymmetry breaking effect is non-local and therefore does not
depend on the ultraviolet completion of the theory. We derived the
correction to the potential using the powerful supersymmetric
approach which only requires to compute the correction to the
K\"ahler potential in the tuned limit of the model.

The model analyzed in this paper shares some of the basic features
of the flux compactifications of string theory which have attracted
a lot of attention recently, starting with \cite{kklt}. In these
constructions one considers a compactification of $10D$ supergravity
on a Calabi-Yau manifold to four dimensions. By adding fluxes,
branes and including non-perturbative effects, it is possible to
stabilize all the moduli of the theory. The ground state is then a
supersymmetric AdS background with the masses of the volume modulus
proportional to the curvature of AdS space. The strategy  for
constructing semi-realistic models is then to start with a vacuum
with large negative cosmological constant, and add
supersymmetry-breaking effects such as anti-$D$ branes that lift the
vacuum energy to a tiny positive value as required by observations.
Under the assumption that this last step does not jeopardize the
stabilization of the scalars, one finally obtains a compactification
with all the moduli stabilized and small positive cosmological
constant.

Let us now cast our results in light of the discussion of the
previous paragraph. The stabilization of the radial modulus is
similar to the one in~\cite{kklt}. In fact the superpotential
responsible for the stabilization of the radion has the same form as
the one in \cite{kklt} but while in our case this superpotential is
induced at the classical level, in \cite{kklt} it is a
non-perturbative effect\footnote{Note however that the K\"ahler
potentials are different at least in the large warping limit.}. Due
to the fact that the ground state is AdS, the masses of the scalars
within the same multiplet are different. This guarantees that in any
AdS compactification at least half of the moduli will have a
potential (of course with the caveat that in AdS the masses of the
scalars could be negative). The main difference in the detuned RS
case is that one of the scalars is massless at tree level but as we
have seen it acquires a mass radiatively. It would be interesting
then to build a toy model of flux compactification based on the
detuned RS model which would allow to test the consistency of the
construction in a simple model.
In some of the examples we considered here, the
negative tree-level potential was indeed modified
by a positive supersymmetry-breaking contribution.
However, the resulting new maximum of the potential was stable
because the total cosmological constant was still
negative, so that the space was AdS$_4$.
Clearly, such a maximum would no longer be stable
if the correction leads to a net zero or positive
cosmological constant.

Finally let us mention that the computation presented in this paper
may be of interest from the point of view of the AdS/CFT
correspondence. We computed a finite loop effect in a weakly coupled
gravity theory with AdS$_4$ background. This corresponds to a
subleading $1/N$ effect in the dual large $N$ three dimensional
conformal field theory. It would be interesting to understand our
results from the point of view of the corresponding $3D$ field
theory.

\section{Acknowledgements}
We thank J.J. Blanco-Pillado, Roberto Contino, Shinji Hirano, Lisa
Randall and Matt Schwartz for discussions. The research of A.K.,
Y.S. and Y.S. is supported by the United States-Israel Science
Foundation (BSF) under grant 2002020. The research of Ya.S. and A.K.
is also supported by the Israel Science Foundation (ISF) under grant
29/03. The work of M.R. is supported by NSF grant PHY-0245068. The
research of Yu.S. is supported by the US Department of Energy under
contract W-7405-ENG-36. M.R., Y.S. and Y.S. would like to thank the
Aspen Center for Physics, where part of this work was completed.
A.K. also thanks LANL for hospitality.



\begin{thebibliography}{nn}
\parskip 0pt

\bibitem{kklt}
  S.~Kachru, R.~Kallosh, A.~Linde and S.~P.~Trivedi,
  Phys.\ Rev.\ D {\bf 68}, 046005 (2003) [hep-th/0301240].

\bibitem{Randall:1999ee}
  L.~Randall and R.~Sundrum,
  Phys.\ Rev.\ Lett.\  {\bf 83}, 3370 (1999), [hep-ph/9905221].

\bibitem{old}
R.~Altendorfer, J.~Bagger and D.~Nemeschansky, ``Supersymmetric
Randall-Sundrum scenario,'' Phys. Rev. D {\bf 63}, 125025 (2001),
\eprint{hep-th/0003117};
T.~Gherghetta and A.~Pomarol, ``Bulk fields and supersymmetry in a
slice of AdS,'' Nucl.\ Phys.\ B {\bf 586}, 141 (2000), \eprint{
  hep-ph/0003129};  A.~Falkowski,
Z.~Lalak and S.~Pokorski, ``Supersymmetrizing branes with bulk in
five-dimensional supergravity,'' Phys.\ Lett.\ B {\bf 491}, 172
(2000), \eprint{hep-th/hep-th/0004093}.

\bibitem{rk}
This is actually the supersymmetric, 2-brane version
of the models studied in:
  N.~Kaloper,
  Phys.\ Rev.\ D {\bf 60}, 123506 (1999),
  [hep-th/9905210],
  A.~Karch and L.~Randall,
  JHEP {\bf 0105}, 008 (2001),
  [hep-th/0011156].

\bibitem{detuned}
J.~Bagger and D.~V.~Belyaev,
Phys.\ Rev.\ D {\bf 67}, 025004 (2003) [hep-th/0206024]; J.~Bagger
and D.~V.~Belyaev,
JHEP 0306:013,2003 [hep-th/0306063].

\bibitem{Bagger:2003dy}
  J.~Bagger and M.~Redi,
  JHEP {\bf 0404}, 031 (2004), [hep-th/0312220].

\bibitem{Bagger:2003fy}
  J.~Bagger and M.~Redi,
  Phys.\ Lett.\ B {\bf 582}, 117 (2004), [hep-th/0310086];
  Z.~Lalak and R.~Matyszkiewicz,
  Phys.\ Lett.\ B {\bf 583}, 364 (2004)
  [hep-th/0310269].

\bibitem{quiros} G.~von Gersdorff, M.~Quiros and A.~Riotto,
Nucl.\ Phys.\ B {\bf 634}, 90 (2002) [hep-th/0204041].

\bibitem{norman}
  J.~P.~Norman,
  Phys.\ Rev.\ D {\bf 69}, 125015 (2004)
  [hep-th/0403298].

\bibitem{cft}
  N.~Arkani-Hamed, M.~Porrati and L.~Randall,
  JHEP {\bf 0108}, 017 (2001)
  [hep-th/0012148];  R.~Rattazzi and A.~Zaffaroni,
  JHEP {\bf 0104}, 021 (2001)
  [hep-th/0012248].

\bibitem{Gregoire:2004nn}
  T.~Gregoire, R.~Rattazzi, C.~A.~Scrucca, A.~Strumia and E.~Trincherini,
  Nucl.\ Phys.\ B {\bf 720}, 3 (2005) [hep-th/0411216].

\bibitem{goldberger}
  W.~D.~Goldberger and I.~Z.~Rothstein,
  Phys.\ Lett.\ B {\bf 491}, 339 (2000), [hep-th/0007065].

\bibitem{Falkowski:2005fm}
  A.~Falkowski,
  JHEP {\bf 0505}, 073 (2005), [hep-th/0502072].

\bibitem{pomarol}
  T.~Gherghetta and A.~Pomarol,
  Nucl.\ Phys.\ B {\bf 586}, 141 (2000)
  [hep-ph/0003129].

\end{thebibliography}
\end{document}